# Non-Abelian Electric-Magnetic Duality with Supersymmetry in 4D and 10D

Hitoshi Nishino[1] and Subhash Rajpoot[2]

*Department of Physics & Astronomy*
*California State University*
*1250 Bellflower Boulevard*
*Long Beach, CA 90840*

**Abstract**

We present electric-magnetic (Hodge) duality formulation for non-Abelian gauge groups with $N = 1$ supersymmetry in $3 + 1$ (4D) dimensions. Our system consists of three multiplets: (i) A super-Yang-Mills vector multiplet (YMVM) $(A_\mu{}^I, \lambda^I)$, (ii) a dual vector multiplet (DVM) $(B_\mu{}^I, \chi^I)$, and (iii) an unphysical tensor multiplet (TM) $(C_{\mu\nu}{}^I, \rho^I, \varphi^I)$, with the index $I$ for adjoint representation. The multiplets YMVM and DVM are dual to each other like: $G_{\mu\nu}{}^I = (1/2)\ \epsilon_{\mu\nu}{}^{\rho\sigma} F_{\rho\sigma}{}^I$. The TM is unphysical, but still plays an important role for establishing the total consistency of the system, based on recently-developed tensor-hierarchy formulation. We also apply this technique to non-Abelian electric-magnetic duality in $9 + 1$ (10D) dimensions. The extra bosonic auxiliary field $K_{\mu_1 \cdots \mu_6}$ in 10D is shown to play an important role for the closure of supersymmetry on fields.

[1] E-Mail: h.nishino@csulb.edu

[2] E-Mail: subhash.rajpoot@csulb.edu

## 1. Introduction

It is conjectured that the discrete group $SL(2, \mathbb{Z}) \subset SL(2, \mathbb{R})$ is the exact symmetry of the full heterotic string theory [1][2], associated with the target-space duality symmetry $SO(6, 22)$ in compactifications to four dimensions (4D). This feature also leads to electric-magnetic (EM) duality in 4D or higher dimensions with lagrangian formulations [3]. The drawback of non-manifest Lorentz invariance in [3] was overcome by the manifestly Lorentz-invariant reformulation [4]. The S-duality between the strong and weak string-couplings is also reduced to EM-duality in 4D [5], making D3-branes self-dual [6].

The $SL(2, \mathbb{R})$ symmetry for a vector field was pointed out early in 1980's [7], and is confirmed to be valid, even in the presence of Dirac-Born-Infeld interactions [7][8]. The $N = 1$ and $N = 2$ supersymmetric generalizations have also been accomplished in [9]. Moreover, this duality-symmetry can be generalized to self-duality in even dimensions [10].

In 4D, the EM-duality is $F_{\mu\nu}{}^I = (1/2)\,\epsilon_{\mu\nu}{}^{\rho\sigma} G_{\rho\sigma}{}^I$, where $G_{\rho\sigma}{}^I$ is the field strength of a new vector field $B_\mu{}^I$ with the adjoint index $I$. However, due to the inconsistency arising for the naïve definition of the field-strength $G^{(0)I}_{\mu\nu} \equiv 2D_{[\mu} B_{\nu]}{}^I$ for a *non-Abelian* vector $B_\mu{}^I$ [11], such an attempt was again bound to fail in the past. This had been the fate of vector fields with non-Abelian indices, not to mention its supersymetrization.

This problem was first solved by the work by Samtleben [12] with the purely bosonic EM-duality for non-Abelian YM gauge field with its Hodge-dual field. The essential ingredient is to introduce Chern-Simons-like terms in the $G$-field strength, combined with a new tensor field $C_{\mu\nu\rho}{}^I$ in the adjoint representation. Subsequently, this result was further generalized in terms of 'tensor-hierarchy formulations' [13][14].

The next natural step is the supersymmetrization of EM-duality for non-Abelian YM gauge fields. Motivated by this viewpoint, we carry out two objectives in this paper: (i) The $N = 1$ supersymmetrization of the system purely-bosonic EM-duality in 4D [12], and (ii) Its generalization to $N = (1, 0)$ YM multiplet in 10D. Even though EM-duality for non-Abelian groups had been known in supergravity, such as $N = 8$ supergravity in 4D with local $SO(8)$, and despite the *purely-bosonic* EM-duality system had been presented as tensor-hierarchy formulation, our new ingredient is the *supersymmetrization* of EM-duality with arbitrary YM groups.

In our formulation in 4D, we introduce the following three multiplets: (i) A super-Yang-Mills vector multiplet (YMVM) which is the conventional vector multiplet, (ii) a dual vector multiplet (DVM) with the field-strength dual to the YM-field-strength, and (iii) a tensor multiplet (TM). The TM plays an important role for the closure of supersymmetry with *no* physical degree of freedom.

The introduction of an extra vector field $B_\mu{}^I$ with the adjoint index in addition to the YM-gauge field $A_\mu{}^I$ is *not* new. In addition to [12], another example is the supersymmetric Jackiw-Pi (JP) model in 3D [15]. The objective of the original JP-model [16] was to improve the parity-odd feature with Chern-Simons (CS) theory in 3D, by introducing an extra vector $B_\mu{}^I$ with the adjoint index. Thus, the introduction of the extra vector $B_\mu{}^I$ is common to our present EM-duality formulation and supersymmetric JP-model [15].

As a by-product of our 4D result, we apply the same mechanism to 10D YM multiplet. The needed field-content is the YMVM $(A_\mu{}^I, \lambda^I)$, DVM $(B_{[7]}{}^I, \chi^I)$[3] and auxiliary tensor potential fields $C_{[8]}{}^I$ and $K_{[6]}$. Here the potentials $A_\mu{}^I$ and $B_{[7]}{}^I$ have respectively the field-strengths $F_{\mu\nu}{}^I$ and $G_{[8]}{}^I$ dual to each other. The important role played by the extra tensor $K_{[6]}$ is explained both in component and superspace languages.

From a certain viewpoint, our formulation seems just a 'trivial' truncation of well-known non-Abelian $N = 1$ systems [17][14][18]. This is because similar structures are found in [17][14][18], after first embedding all fields in super-multiplets and then truncating out all extra fields. Conceptually, that is one way to describe our objective. In practice, however, the most non-trivial process is the realization of such 'truncation' consistently with supersymmetry. Whereas the purely-bosonic part of our system had been presented in [12], its supersymmetrization is the most non-trivial part. As we will see also, the necessity of the auxiliary tensor $K_{[6]}$ in the 10D case characterizes our non-trivial formulation.

Our paper is organized as follows: In the next section, we review the tensor-hierarchy formulation [13][14] applied to EM-duality. In section 3, we give the $N = 1$ supersymmetrization of *non-Abelian* EM-duality. In section 4, we re-formulate our theory in terms of superspace language [19]. We next apply the 4D result to 10D super YM multiplet in component in sec. 5. In sec. 6, we present its superspace re-formulation. Concluding remarks are given in section 7.

## 2. Tensor-Hierarchy and Duality

Our field content consists of three multiplets: (i) A YMVM: $(A_\mu{}^I, \lambda^I)$, (ii) a DVM: $(B_\mu{}^I, \chi^I)$, and (iii) a TM: $(C_{\mu\nu}{}^I, \rho^I, \varphi^I)$. The vector fields $A_\mu{}^I$, $B_\mu{}^I$, and the tensor field

[3] We use the symbol $[n]$ like $X_{[n]} \equiv X_{\mu_1 \cdots \mu_n}$ to save space for indices.

$C_{\mu\nu}{}^I$ have the following field-strengths defined by [12][13][14]

$$F_{\mu\nu}{}^I \equiv + 2\partial_{[\mu}A_{\nu]}{}^I + mf^{IJK}A_\mu{}^J A_\nu{}^K \quad , \tag{2.1a}$$

$$G_{\mu\nu}{}^I \equiv + 2D_{[\mu}B_{\nu]}{}^I + mC_{\mu\nu}{}^I \equiv +2\left(\partial_{[\mu}B_{\nu]}{}^I + mf^{IJK}A_{[\mu}B_{\nu]}{}^I\right) + mC_{\mu\nu}{}^I \quad , \tag{2.1b}$$

$$H_{\mu\nu\rho}{}^I \equiv = +3D_{[\mu}C_{\nu\rho]}{}^I + 3f^{IJK}F_{\mu\nu}{}^J B_{\rho]}{}^K \quad . \tag{2.1c}$$

We use $m$ as the YM-gauge coupling constant. These structures with the Chern-Simons (CS) like-terms in $G$ and $H$-field-strengths follow the general pattern in the recently-developed tensor-hierarchy formulations [13][14]. Accordingly, the field-strengths $F$, $G$ and $H$ satisfy their proper Bianchi-identities (BIds):

$$D_{[\mu}F_{\nu\rho]}{}^I \equiv 0 \quad , \tag{2.2a}$$

$$D_{[\mu}G_{\nu\rho]}{}^I \equiv + \frac{1}{3}\, mH_{\mu\nu\rho}{}^I \quad , \tag{2.2b}$$

$$D_{[\mu}H_{\nu\rho\sigma]}{}^I \equiv + \frac{3}{2}\, f^{IJK}F_{[\mu\nu}{}^J G_{\rho\sigma]}{}^K \quad . \tag{2.2c}$$

The general variation of these field-strengths are given by

$$\delta F_{\mu\nu}{}^I = + 2D_{[\mu|}(\delta A_{|\nu]}{}^I) \quad , \tag{2.3a}$$

$$\delta G_{\mu\nu}{}^I = + 2D_{[\mu|}(\delta B_{|\nu]}{}^I) + m(\widetilde{\delta} C_{\mu\nu}{}^I) \quad , \tag{2.3b}$$

$$\delta H_{\mu\nu\rho}{}^I = + 3D_{[m||}\left(\widetilde{\delta} C_{|\nu\rho]}{}^I\right) - 3f^{IJK}(\delta B_{[\mu|}{}^J)F_{|\nu\rho]}{}^K + 3f^{IJK}(\delta A_{[\mu|}{}^J)\, G_{|\nu\rho]}{}^K \quad , \tag{2.3c}$$

$$\widetilde{\delta} C_{\mu\nu}{}^I \equiv \delta C_{\mu\nu}{}^I + 2f^{IJK}(\delta A_{[\mu|}{}^J)B_{|\nu]}{}^I \quad . \tag{2.3d}$$

Since the dual-vector $B_\mu{}^I$ has a space-time index $\mu$, it must have its proper 'gauge' transformation: $\delta_U B_\mu{}^I = D_\mu \beta^I$. The tensor $C_{\mu\nu}{}^I$ should also have its tensorial gauge transformation: $\delta_V C_{\mu\nu}{}^I = 2D_{[\mu}\gamma_{\nu]}{}^I$ [13][14]. In total, there are three different (generalized) gauge and tensor transformations $\delta_T$, $\delta_U$ and $\delta_V$ with the appropriate parameters $\alpha^I$, $\beta^I$ and $\gamma_\mu{}^I$ [13][14]:

$$\delta_T(A_\mu{}^I,\ B_\mu{}^I,\ C_{\mu\nu}{}^I) = (D_\mu\alpha^I,\ - mf^{IJK}\alpha^J B_\mu{}^K,\ - mf^{IJK}\alpha^J B_\mu{}^K) \quad , \tag{2.4a}$$

$$\delta_U(A_\mu{}^I,\ B_\mu{}^I,\ C_{\mu\nu}{}^I) = (0,\ + D_\mu\beta^I,\ + f^{IJK}\beta^J F_{\mu\nu}{}^K) \quad , \tag{2.4b}$$

$$\delta_V(A_\mu{}^I,\ B_\mu{}^I,\ C_{\mu\nu}{}^I) = (0,\ - m\gamma_\mu{}^I,\ + 2D_{[\mu}\gamma_{\nu]}{}^I) \quad . \tag{2.4c}$$

Using (2.4) in (2.3), we get

$$\delta_T(F_{\mu\nu}{}^I,\ G_{\mu\nu}{}^I,\ H_{\mu\nu\rho}{}^I) = - mf^{IJK}\alpha^J(F_{\mu\nu}{}^K,\ G_{\mu\nu}{}^K,\ H_{\mu\nu\rho}{}^K) \quad , \tag{2.5a}$$

$$\delta_U(F_{\mu\nu}{}^I,\ G_{\mu\nu}{}^I,\ H_{\mu\nu\rho}{}^I) = (0,\ 0,\ 0) \quad , \qquad \delta_V(F_{\mu\nu}{}^I,\ G_{\mu\nu}{}^I,\ H_{\mu\nu\rho}{}^I) = (0,\ 0,\ 0) \quad . \tag{2.5b}$$

In particular, the CS-like terms in the $G$ and $H$-field-strengths play important roles for the $\delta_U$ and $\delta_V$-invariances (2.5b). These results simply follow from the straightforward application of the more general tensor-hierarchy formulation [13][14].

Our crucial starting point is to require the EM-duality between the field-strengths $F$ and $G$:[4)]

$$G_{\mu\nu}{}^I \overset{*}{=} + \frac{1}{2}\,\epsilon_{\mu\nu}{}^{\rho\sigma} F_{\rho\sigma}{}^I \quad . \tag{2.6}$$

The relative sign between these two equations is negative, because of our metric signature $(+,+,+,-)$. Note that the RHS of the $H$-BI (2.2c) vanishes upon the use of the EM-duality (2.6).

Before the discovery of tensor-hierarchy formulation [13][14], there used to exist inconsistency for EM-duality for *non-Abelian* groups. For example, the gauge non-covariance is one of them. The naïvely-defined field-strength

$$G_{\mu\nu}^{(0)I} \equiv +2 D_{[\mu} B_{\nu]}{}^I \tag{2.7}$$

is *not* $\delta_\beta$-invariant, because it transforms as

$$\delta_U G_{\mu\nu}^{(0)I} = m f^{IJK} F_{\mu\nu}{}^J \beta^K \neq 0 \quad . \tag{2.8}$$

The trouble is that this transformation does *not* leave the duality condition (2.6) intact. What is needed is an extra term in $G_{\mu\nu}{}^I$ as in (2.1b) that cancels the unwanted term (2.8), yielding $\delta_U G_{\mu\nu}{}^I = 0$. In contrast, the *non*-invariance of the naïve field-strength $\delta_U G_{\mu\nu}^{(0)I} \neq 0$ used to present an obstruction to establish the EM-duality: $G_{\mu\nu}^{(0)I} \overset{*}{=} (1/2)\,\epsilon_{\mu\nu}{}^{\rho\sigma} F_{\rho\sigma}{}^I$.

## 3. Supersymmetric EM-Duality

The next step is to supersymmetrize the duality condition (2.6). Because of the general tensor-hierarchy formulation [13], this process is straightforward. As has been mentioned, the TM in our system is *unphysical*, namely, all fields $(C_{\mu\nu}{}^I, \rho^I, \varphi^I)$ have *no* physical degree of freedom.

To be more specific, the $N=1$ supersymmetry transformation rule for our multplets YMVM, DVM and TM is

$$\delta_Q A_\mu{}^I = + (\overline{\epsilon} \gamma_\mu \lambda^I) \quad , \tag{3.1a}$$

[4)] We use the symbol $\overset{*}{=}$ for an equality related to a duality, or a more general constraint related to consistency with duality. Similarly, we use the symbol $\doteq$ for a field equation.

$$\delta_Q \lambda^I = + \frac{1}{2}(\gamma^{\mu\nu}\epsilon) F_{\mu\nu}{}^I + im(\gamma_5\epsilon)\varphi^I \quad , \tag{3.1b}$$

$$\delta_Q B_\mu{}^I = + i(\overline{\epsilon}\gamma_5\gamma_\mu\chi^I) \quad , \tag{3.1c}$$

$$\delta_Q \chi^I = + \frac{i}{2}(\gamma_5\gamma^{\mu\nu}\epsilon) G_{\mu\nu}{}^I - im(\gamma_5\epsilon)\varphi^I \tag{3.1d}$$

$$\delta_Q C_{\mu\nu}{}^I = + i(\overline{\epsilon}\gamma_5\gamma_{\mu\nu}\rho^I) - 2f^{IJK}(\overline{\epsilon}\gamma_{[\mu|}\lambda^J) B_{|\nu]}{}^K \quad , \tag{3.1e}$$

$$\delta_Q \rho^I = - \frac{i}{6}(\gamma_5\gamma^{\mu\nu\rho}\epsilon)\, H_{\mu\nu\rho}{}^I + i(\gamma_5\gamma^\mu\epsilon) D_\mu\varphi^I + \frac{1}{2} f^{IJK}(\gamma_\mu\epsilon)(\overline{\lambda}{}^J\gamma^\mu\gamma^K) \quad , \tag{3.1f}$$

$$\delta_Q \varphi^I = + i(\overline{\epsilon}\gamma_5\rho^I) \quad . \tag{3.1g}$$

Accordingly, by the use of (2.3) we can get

$$\delta_Q F_{\mu\nu}{}^I = - 2(\overline{\epsilon}\gamma_{[\mu} D_{\nu]}\lambda^I) \quad , \qquad \delta_Q G_{\mu\nu}{}^I = -2i(\overline{\epsilon}\gamma_5\gamma_{[\mu} D_{\nu]}\chi^I) + im(\overline{\epsilon}\gamma_5\gamma_{\mu\nu}\rho^I) \quad , \tag{3.2a}$$

$$\delta_Q H_{\mu\nu\rho}{}^I = + 3i(\overline{\epsilon}\gamma_5\gamma_{[\mu\nu} D_{\rho]}\rho^I) + 3f^{IJK}(\overline{\epsilon}\gamma_{[\mu|}\lambda^J) G_{|\nu\rho]}{}^K - 3if^{IJK}(\overline{\epsilon}\gamma_5\gamma_{[\mu}\chi^J) F_{\nu\rho]}{}^K \quad . \tag{3.2b}$$

The definitions for the $F$, $G$ and $H$-field-strengths are exactly the same as in (2.1).

Our supersymmetric completion of the duality (2.6) reads as

$$G_{\mu\nu}{}^I \overset{*}{=} + \frac{1}{2}\epsilon_{\mu\nu}{}^{\rho\sigma} F_{\rho\sigma}{}^I \quad , \tag{3.3a}$$

$$\lambda^I \overset{*}{=} - \chi^I \quad , \qquad \not{D}\lambda^I \doteq 0 \quad , \qquad \not{D}\chi^I \doteq 0 \quad , \tag{3.3b}$$

$$\rho^I \overset{*}{=} 0 \quad , \qquad \varphi^I \overset{*}{=} 0 \quad , \tag{3.3c}$$

$$H_{\mu\nu\rho}{}^I \overset{*}{=} - \frac{i}{2} f^{IJK}(\overline{\lambda}{}^J\gamma_5\gamma_{\mu\nu\rho}\lambda^K) \quad . \tag{3.3d}$$

Some remarks are in order: First, the last two equations in (3.3b) are actually field equations, but they are still indirectly related to the EM-duality by supersymmetry. Second, the first equation in (3.3b) implies that the two fermions $\lambda$ and $\chi$ coincide up to a sign. Third, (3.3c) is needed, so that the TM is *not* physical. Fourth, the condition on $H$ is non-trivial, because if we simply put $H_{\mu\nu\rho}{}^I \overset{*}{=} 0$, then its supersymmetric transformation generates *non*-vanishing terms on-shell due to (3.2b). Even though the first term in (3.2b) vanishes due to (3.3c), the additional two terms $\approx (\overline{\epsilon}\gamma_5\gamma\lambda)\wedge G$ and $(\overline{\epsilon}\gamma\chi)\wedge F$ remain. Even though the latter is *approximately* equivalent to the former because of (3.3a) and (3.3b), they do *not exactly* cancel each other. It is the variation of the RHS of (3.3d) that cancels these two terms: $\delta_Q[\, H_{\mu\nu\rho}{}^I + (i/2) f^{IJK}(\overline{\lambda}{}^J\gamma_5\gamma_{\mu\nu\rho}\lambda^K)\,] \overset{*}{=} 0$.

Fifth, all other equations in (3.3) are consistent with supersymmetry. This confirms the total on-shell consistency with supersymmetry.

Sixth, the closure of supersymmetry works as follows:

$$
\begin{aligned}
&[\delta_{Q_1}, \delta_{Q_2}] = \delta_{P_3} + \delta_{T_3} + \delta_{U_3} + \delta_{V_3} \ , \\
&\xi_3^\mu = +2(\overline\epsilon_1 \gamma^\mu \epsilon_2) \ , \quad \alpha_3^I = -\xi_3^\mu A_\mu{}^I \ , \quad \beta_3^I = -\xi_3^\mu B_\mu{}^I \ , \quad \gamma_{3\mu}{}^I = -\xi_3^\nu C_{\nu\mu}{}^I - \xi_{3\,\mu} \varphi^I \ , \qquad (3.4)
\end{aligned}
$$

where $\delta_P$ is the translation operation. The transformations $\delta_P$, $\delta_T$, $\delta_U$ and $\delta_V$ respectively have the parameters $\xi^\mu$, $\alpha^I$, $\beta^I$ and $\gamma_\mu{}^I$. The subscript $_3$ on these parameters is to show that they are produced out of the commutator $[\delta_{Q_1}, \, \delta_{Q_2}]$.

Seventh, other commutators among $\delta_U$ and $\delta_Q$ or $\delta_V$ and $\delta_Q$ are the following:

$$
[\delta_Q, \delta_U] = \delta_V \ , \qquad \gamma_\mu{}^I \equiv - f^{IJK} (\overline\epsilon \gamma_\mu \lambda^J) \beta^K \ , \qquad (3.5\text{a})
$$

$$
[\delta_{T_1}, \delta_{T_2}] = \delta_{T_3} \ , \qquad \alpha_3^I \equiv - f^{IJK} \alpha_1^J \alpha_2^K \ , \qquad (3.5\text{b})
$$

$$
\begin{aligned}
[\delta_Q, \delta_T] &= [\delta_Q, \delta_V] = [\delta_T, \delta_U] = [\delta_T, \delta_V] = [\delta_U, \delta_V] = [\delta_{U_1}, \delta_{U_2}] = [\delta_{V_1}, \delta_{V_2}] = [\delta_{K_1}, \delta_{K_2}] \\
&= [\delta_T, \delta_K] = [\delta_U, \delta_K] = [\delta_V, \delta_K] = 0 \ . \qquad (3.5\text{c})
\end{aligned}
$$

Eighth, the degrees of freedom (DOF) in our system are counted as follows: The TM is off-shell without auxiliary fields. However, since it is *unphysical* with DOF are $0+0$ on-shell. Both of our YMVM and DVM are *on-shell*, namely, there is *no* $D$-type auxiliary fields. So the total DOF of these two multiplets are $2(2+2)$ on-shell. However, due to the supersymmetric duality (3.3a) and (3.3b), the total DOF are reduced to $2(2+2)/2 = 2+2$.

This situation is very similar to the duality-symmetric 11D supergravity [20]. Namely, in [20] we use both the 4-th rank field-strength $F_{\mu\nu\rho\sigma}$ and its Hodge dual $G_{\mu_1 \cdots \mu_7}$ simultaneously. Originally, there are $2 \binom{9}{3} = 2 \cdot 84 = 168$ on-shell DOF, but due to the duality relation $F_{[4]} = (1/7!) \, \epsilon_{[4]}{}^{[7]} G_{[7]}$, the total DOF are reduced again to 84, balancing the usual $128+128$ on-shell DOF in 11D supergravity [21].

## 4. Superspace Re-Formulation

Once we have established the component formulation of our system, it is rather straightforward to translate it into superspace [19]. Our superfield-strengths are $F_{AB}{}^I$, $G_{AB}{}^I$ and $H_{ABC}{}^I$,[5] defined by

$$
F_{AB}{}^I \equiv \ + E_{[A} A_{B)}{}^I - T_{AB}{}^C A_C{}^I + m f^{IJK} A_A{}^J A_B{}^K \ , \qquad (4.1\text{a})
$$

[5] In superspace, we use the local coordinate indices $A \equiv (a,\alpha)$, $B \equiv (b,\beta)$, $\cdots$ for the bosonic (or fermionic) coordinates $a, \, b, \, \cdots = 0, \, 1, \, 2, \, 3$ (or $\alpha, \, \beta, \, \cdots = 1, \, 2, \, 3, \, 4$). The (anti)symmetrization in superspace is such as $M_{[AB)} \equiv M_{AB} - (-)^{AB} M_{BA}$. The YM-covariant derivative $D_\mu$ in component language is now $\nabla_a$. For curved coordinates, we use $M, \, N, \, \cdots$.

$$G_{AB}{}^I \equiv + \nabla_{[A} B_{B)}{}^I - T_{AB}{}^C B_C{}^I + m\, C_{AB}{}^I \quad , \tag{4.1b}$$

$$H_{ABC}{}^I \equiv + \frac12 \nabla_{[A} C_{BC)}{}^I - \frac12 T_{[AB|}{}^D C_{D|C)}{}^I + \frac12 f^{IJK} F_{[AB}{}^J B_{C)}{}^K \quad , \tag{4.1c}$$

where $E_A \equiv E_A{}^M \partial_M$, while $\nabla_A$ is the YM-gauge covariant derivative: $\nabla_A \equiv E_A{}^M \partial_M + A_M{}^I \tau^I$ with the YM group generators $\tau^I$. These field-strengths satisfy their respective BIds:

$$+ \frac12 \nabla_{[A} F_{BC)}{}^I - \frac12 T_{[AB|}{}^D F_{D|C)}{}^I \equiv 0 \quad , \tag{4.2a}$$

$$+ \frac12 \nabla_{[A} G_{BC)}{}^I - \frac12 T_{[AB|}{}^D G_{D|C)}{}^I - m H_{ABC}{}^I \equiv 0 \quad , \tag{4.2b}$$

$$+ \frac16 \nabla_{[A} H_{BCD)}{}^I - \frac14 T_{[AB|}{}^E H_{E|CD)}{}^I - \frac14 f^{IJK} F_{[AB|}{}^J G_{|CD)}{}^K \equiv 0 \quad . \tag{4.2c}$$

Eqs. (4.1) and (4.2) are nothing but our component results (2.1) and (2.2) re-casted into superspace [19].

Our superspace constraints at engineering dimensions $0 \le d \le 1$[6] are

$$T_{\alpha\beta}{}^c = +2(\gamma^c)_{\alpha\beta} \quad , \qquad H_{\alpha\beta c}{}^I = +2(\gamma_c)_{\alpha\beta}\, \varphi^I \quad , \tag{4.3a}$$

$$F_{\alpha b}{}^I = -(\gamma_b \lambda^I)_\alpha \quad , \qquad G_{\alpha b}{}^I = -i(\gamma_5 \gamma_b \chi^I)_\alpha \quad , \qquad H_{\alpha bc}{}^I = -i(\gamma_5 \gamma_{bc} \rho^I)_\alpha \quad , \tag{4.3b}$$

$$\nabla_\alpha \lambda_\beta{}^I = + \frac12 (\gamma^{cd})_{\alpha\beta} F_{cd}{}^I - im(\gamma_5)_{\alpha\beta}\, \varphi^I \quad , \tag{4.3c}$$

$$\nabla_\alpha \chi_\beta{}^I = + \frac{i}{2} (\gamma_5 \gamma^{cd})_{\alpha\beta} G_{cd}{}^I + im(\gamma_5)_{\alpha\beta}\, \varphi^I \quad , \tag{4.3d}$$

$$\nabla_\alpha \rho_\beta{}^I = - \frac{i}{6} (\gamma_5 \gamma^{cde})_{\alpha\beta} H_{cde}{}^I - i(\gamma_5 \gamma^c)_{\alpha\beta} \nabla_c \varphi^I + \frac12 f^{IJK} (\gamma^c)_{\alpha\beta} (\overline\lambda{}^J \gamma_c \lambda^K) \quad , \tag{4.3e}$$

$$\nabla_\alpha \varphi^I = -i(\gamma_5 \rho^I)_\alpha \quad . \tag{4.3f}$$

Other independent components, such as $H_{\alpha\beta\gamma}{}^I$ are all zero.

The constraints at $d = 3/2$ are equivalent to (3.2):

$$\nabla_\alpha F_{bc}{}^I = (\gamma_{[b} \nabla_{c]} \lambda^I)_\alpha \quad , \qquad \nabla_\alpha G_{bc}{}^I = i(\gamma_5 \gamma_{[b} \nabla_{c]} \chi)_\alpha - im(\gamma_5 \gamma_{bc} \rho^I)_\alpha \quad , \tag{4.4a}$$

$$\nabla_\alpha H_{bcd}{}^I = - \frac{i}{2} (\gamma_5 \gamma_{[bc} \nabla_{d]} \rho^I)_\alpha - \frac12 f^{IJK} (\gamma_{[b} \lambda^J)_\alpha G_{cd]}{}^K + \frac{i}{2} f^{IJK} (\gamma_{[b} \chi^J)_\alpha F_{cd]}{}^K \quad . \tag{4.4b}$$

Our duality-related equations in (3.3) are re-expressed as

$$G_{ab}{}^I \overset{*}{=} + \frac12 \epsilon_{ab}{}^{cd} F_{cd}{}^I \quad , \tag{4.5a}$$

$$\lambda_\alpha{}^I \overset{*}{=} - \chi_\alpha{}^I \quad , \qquad (\nabla\!\!\!\!/\, \lambda^I)_\alpha \doteq 0 \quad , \qquad (\nabla\!\!\!\!/\, \chi^I)_\alpha \doteq 0 \quad , \tag{4.5b}$$

$$\rho_\alpha{}^I \overset{*}{=} 0 \quad , \qquad \varphi^I \overset{*}{=} 0 \quad , \tag{4.5c}$$

$$H_{abc}{}^I \overset{*}{=} - \frac{i}{2} f^{IJK} (\overline\lambda{}^J \gamma_5 \gamma_{abc} \lambda^K) \quad , \qquad \widetilde H_\mu{}^I \overset{*}{=} + \frac12 f^{IJK} (\overline\lambda{}^J \gamma_\mu \lambda^K) \quad , \tag{4.5d}$$

[6] The engineering dimension for our bosonic (or fermionic) fundamental field is $d = 0$ (or $d = 1/2$).

It is not too difficult to confirm the mutual consistency of these equations. For example, a spinorial derivative $\nabla_\alpha$ on (4.5a) is shown to vanish:

$$\nabla_\alpha \left( G_{ab}{}^I - \frac{1}{2} \epsilon_{ab}{}^{cd} F_{cd}{}^I \right)$$
$$\overset{*}{=} \left( \gamma_{[a} \nabla_{b]} \lambda^I \right)_\alpha + (\gamma_{ab}{}^c \nabla_c \lambda^I)_\alpha \overset{*}{=} (\gamma_{ab} \nabla\!\!\!\!/\, \lambda^I)_\alpha \doteq 0 \quad , \tag{4.6}$$

by the use of (4.5b) and (4.5c).

## 5. 10D Application

As we have promised, we apply our supersymmetrization technique in 4D to 10D super YM system. Our field content is the YMVM $(A_\mu{}^I, \lambda^I)$, DVM $(B_{[7]}{}^I, \chi^I)$, and auxiliary bosonic tensor fields $C_{[8]}{}^I$ and $K_{[6]}$. Here the fermions $\lambda^I$ and $\chi^I$ are both Majorana-Weyl spinors with the positive chirality, as in the conventional super YM theory in 10D. Compared with the previous 4D case, the tensor $K_{[6]}$ is new, without any adjoint index. The important role played by this tensor will be clarified after (5.7c) below.

The $N = (1, 0)$ supersymmetry transformation rule is

$$\delta_Q A_\mu{}^I = + (\overline{\epsilon} \gamma_\mu \lambda^I) \quad , \tag{5.1a}$$

$$\delta_Q \lambda^I = + \frac{1}{2} (\gamma^{\mu\nu} \epsilon) F_{\mu\nu}{}^I \quad , \tag{5.1b}$$

$$\delta_Q B_{\mu_1 \cdots \mu_7}{}^I = + (\overline{\epsilon} \gamma_{\mu_1 \cdots \mu_7} \chi^I) + 7 K_{[\mu_1 \cdots \mu_6|} (\overline{\epsilon} \gamma_{|\mu_7]} \lambda^I) \quad , \tag{5.1c}$$

$$\delta_Q \chi^I = - \frac{1}{8!} (\gamma^{[8]} \epsilon) G_{[8]}{}^I \quad , \tag{5.1d}$$

$$\delta_Q C_{\mu_1 \cdots \mu_8}{}^I = - 8 f^{IJK} (\overline{\epsilon} \gamma_{\mu_1} \lambda^J) B_{\mu_2 \cdots \mu_8]}{}^K \quad , \tag{5.1e}$$

$$\delta_Q K_{\mu_1 \cdots \mu_6} = 0 \quad , \tag{5.1f}$$

where $\gamma_{11} \epsilon = + \epsilon$. The field-strengths $F$, $G$, $H$ and $L$ respectively of the potentials $A$, $B$, $C$ and $K$ are defined by

$$F_{\mu\nu}{}^I \equiv + 2 \partial_{[\mu} A_{\nu]}{}^I + m f^{IJK} A_\mu{}^J A_\nu{}^K \quad , \tag{5.2a}$$

$$G_{\mu_1 \cdots \mu_8}{}^I \equiv + 8 \partial_{[\mu_1} B_{\mu_2 \cdots \mu_8]}{}^I + m C_{\mu_1 \cdots \mu_8}{}^I - 28 K_{[\mu_1 \cdots \mu_6} F_{\mu_7 \mu_8]}{}^I \quad , \tag{5.2b}$$

$$H_{\mu_1 \cdots \mu_9}{}^I \equiv + 9 D_{[\mu_1} C_{\mu_2 \cdots \mu_9]}{}^I + 36 f^{IJK} F_{[\mu_1 \mu_2}{}^J B_{\mu_3 \cdots \mu_9]}{}^K \quad , \tag{5.2c}$$

$$L_{\mu_1 \cdots \mu_7} \equiv + 7 \partial_{[\mu_1} K_{\mu_2 \cdots \mu_7]} \quad . \tag{5.2d}$$

These field-strengths satisfy the BIds

$$D_{[\mu} F_{\nu\rho]}{}^I \equiv 0 \quad , \tag{5.3a}$$

$$D_{[\mu_1} G_{\mu_2\cdots\mu_9]}{}^I \equiv + \frac{1}{9} m H_{\mu_1\cdots\mu_9}{}^I - 4 L_{[\mu_1\cdots\mu_7} F_{\mu_8\mu_9]}{}^I \quad , \tag{5.3b}$$

$$D_{[\mu_1} H_{\mu_2\cdots\mu_{10}]}{}^I \equiv + \frac{9}{2} f^{IJK} F_{[\mu_1\mu_2}{}^J G_{\mu_3\cdots\mu_{10}]}{}^K \quad , \tag{5.3c}$$

$$\partial_{[\mu_1} L_{\mu_2\cdots\mu_8]} \equiv 0 \quad . \tag{5.3d}$$

The arbitrary variations of these field-strengths are

$$\delta F_{\mu\nu}{}^I = + 2 D_{[\mu} (\delta A_{\nu]}{}^I) \quad , \tag{5.4a}$$

$$\begin{aligned} \delta G_{\mu_1\cdots\mu_8}{}^I = & + 8 D_{[\mu_1} (\widetilde\delta B_{\mu_2\cdots\mu_8]}{}^I) - 8 (\delta A_{[\mu_1}{}^I) L_{\mu_2\cdots\mu_8]} \\ & + m (\widetilde\delta C_{\mu_1\cdots\mu_8}) - 28 (\delta K_{[\mu_1\cdots\mu_4}) F_{\mu_7\mu_8]}{}^I \quad , \end{aligned} \tag{5.4b}$$

$$\begin{aligned} \delta H_{\mu_1\cdots\mu_9}{}^I = & + 9 D_{[\mu_1} (\widetilde\delta C_{\mu_2\cdots\mu_9]}{}^I) - 36 f^{IJK} (\widetilde\delta B_{[\mu_1\cdots\mu_7}{}^J) F_{\mu_8\mu_9]}{}^K \\ & + 9 f^{IJK} (\widetilde\delta A_{[\mu_1}{}^J) G_{\mu_2\cdots\mu_9]}{}^K \quad , \end{aligned} \tag{5.4c}$$

$$\delta L_{\mu_1\cdots\mu_7}{}^I = + 7 \partial_{[\mu_1} (\delta K_{\mu_1\cdots\mu_7]}{}^I) \quad , \tag{5.4d}$$

$$\widetilde\delta B_{\mu_1\cdots\mu_7}{}^I \equiv \delta B_{\mu_1\cdots\mu_7}{}^I - 7 (\delta A_{[\mu_1|}{}^I) K_{|\mu_2\cdots\mu_7]} \quad , \tag{5.4e}$$

$$\widetilde\delta C_{\mu_1\cdots\mu_8}{}^I \equiv \delta C_{\mu_1\cdots\mu_8}{}^I + 8 f^{IJK} (\delta A_{[\mu_1}{}^J) B_{\mu_2\cdots\mu_8]}{}^K \quad . \tag{5.4f}$$

There are four different gauge transformations $\delta_T, \ \delta_U, \ \delta_V$ and $\delta_K$:

$$\delta_T A_\mu{}^I = D_\mu \alpha^I \quad , \quad \delta_T (B_{[7]}{}^I, \, C_{[8]}{}^I, \, K_{[6]}) = - m f^{IJK} \alpha^J (B_{[7]}{}^K, \, C_{[8]}{}^K, \, 0) \quad , \tag{5.5a}$$

$$\delta_U B_{\mu_1\cdots\mu_7}{}^I = + 7 D_{[\mu_1} \beta_{\mu_2\cdots\mu_7]} \quad , \quad \delta_U C_{\mu_1\cdots\mu_8}{}^I = - 28 f^{IJK} F_{[\mu_1\mu_2}{}^J \beta_{\mu_3\cdots\mu_8]}{}^K \quad , \tag{5.5b}$$

$$\delta_V B_{[7]}{}^I = - m \gamma_{[7]}{}^I \quad , \quad \delta_V C_{\mu_1\cdots\mu_8}{}^I = + 8 D_{[\mu_1} \gamma_{\mu_2\cdots\mu_8]}{}^I \quad , \tag{5.5c}$$

$$\delta_K B_{\mu_1\cdots\mu_7}{}^I = 21 \kappa_{[\mu_1\cdots\mu_5} F_{\mu_6\mu_7]}{}^I \quad , \quad \delta_K K_{\mu_1\cdots\mu_6} = + 6 \, \partial_{[\mu_1} \kappa_{\mu_2\cdots\mu_6]} \quad . \tag{5.5d}$$

for the potentials $A$, $B$, $C$ and $K$, respectively. All other fields not given above are *invariant*, *e.g.*, $\delta_U A_\mu{}^I = 0$, or $\delta_V K_{[6]} = 0$. Under each of $\delta_U$, $\delta_V$ and $\delta_K$-transformations, there are only two fields transforming. Note that $B_{[7]}{}^I$ also transforms under $\delta_K$. Using (5.4), we can prove the covariance and invariance of our field-strengths:

$$\delta_T (F_\mu{}^I, G_{[8]}{}^I, H_{[9]}{}^I, L_{[7]}) = - m f^{IJK} \alpha^J (F_\mu{}^K, G_{[8]}{}^K, H_{[9]}{}^K, 0) \quad ,$$

$$\delta_U (F_\mu{}^I, G_{[8]}{}^I, H_{[9]}{}^I, L_{[7]}) = (0,0,0,0) \quad , \quad \delta_U (F_\mu{}^I, G_{[8]}{}^I, H_{[9]}{}^I, L_{[7]}) = (0,0,0,0) \quad , \tag{5.6a}$$

$$\delta_K (F_\mu{}^I, G_{[8]}{}^I, H_{[9]}{}^I, L_{[7]}) = (0,0,0,0) \quad . \tag{5.6b}$$

The closure of supersymmetry is

$$[\delta_{Q_1}, \delta_{Q_2}] = \delta_{P_3} + \delta_{T_3} + \delta_{U_3} + \delta_{V_3} + \delta_{K_3} \quad , \tag{5.7a}$$

$$\xi_3^\mu \equiv +2(\bar\epsilon_1\gamma^\mu\epsilon_2) \quad , \quad \alpha_3^I \equiv -\xi_3^\mu A_\mu{}^I \quad , \quad \beta_{3\,\mu_1\cdots\mu_6}{}^I \equiv -\xi_3^\mu B_{\nu\mu_1\cdots\mu_6}{}^I \quad , \tag{5.7b}$$

$$\gamma_{3\,\mu_1\cdots\mu_7}{}^I \equiv -\xi_3^\nu C_{\nu\mu_1\cdots\mu_7}{}^I \quad , \quad \kappa_{3\,\mu_1\cdots\mu_5} \equiv +2(\bar\epsilon_1\gamma_{\mu_1\cdots\mu_5}\epsilon_2) - \xi_3^\nu K_{\nu\mu_1\cdots\mu_5} \quad . \tag{5.7c}$$

The closures on $B_{[7]}{}^I$ and $C_{[8]}{}^I$ need special care. In $[\delta_{Q_1}, \delta_{Q_2}]\, B_{[7]}{}^I$, there arises a term $42(\bar\epsilon_1\gamma_{[\mu_1\cdots\mu_5|}\epsilon_2)F_{|\mu_6\mu_7]}{}^I$. Usually, such a term poses a problem, because a $\gamma_{[5]}$-term is *not* acceptable in a supersymmetry-commutator. Even though its leading gradient-term $84(\bar\epsilon_1\gamma_{[\mu_1\cdots\mu_5|}\epsilon_2)\,\partial_{|\mu_6|}A_{|\mu_7]}{}^I$ may be absorbed into $\delta_U B_{[7]}{}^I$, the *non-Abelian* term $42m(\bar\epsilon_1\gamma_{[\mu_1\cdots\mu_5|}\epsilon_2)f^{IJK}A_{|\mu_6}{}^J A_{\mu_7]}{}^K$ can *not* be interpreted as a part of $\delta_U B_{[7]}{}^I$. However, in our system, this problematic term can be interpreted as a $\delta_K$-transformation as $\delta_K B_{\mu_1\cdots\mu_7}{}^I = 21\kappa_{[\mu_1\cdots\mu_5|}F_{|\mu_6\mu_7]}{}^I$ as in (5.5d) and (5.7c). This justifies the necessity of the new gauge symmetry $\delta_K$ for the new field $K_{[6]}$. Note also that in the previous 4D case, the analog of the $K_{[6]}$-field was *not* needed, because there was *no* higher-rank gamma-term in $[\delta_{Q_1}, \delta_{Q_2}]\, B_{\mu\nu}{}^I$, such as $42m(\bar\epsilon_1\gamma_{[\mu_1\cdots\mu_5|}\epsilon_2)f^{IJK}A_{|\mu_6}{}^J A_{\mu_7]}{}^K$. This is the very reason why we need $K_{[6]}$ in 10D with its associated symmetry $\delta_K$. The necessity of $K_{[6]}$ is also reflected in superspace language [19] in the next section.

Some readers may still wonder what is the real role played by the tensor $K_{[6]}$. Such a question seems legitimate, because the field strength $L_{[7]}$ is zero, so $K_{[6]}$ is unphysical, and completely gauged away. This question is answered as follows: If $K_{[6]}$ *were* gauged away, and its gauge transformation $\delta_K$ *were* no longer available, the aforementioned unwanted term $42m(\bar\epsilon_1\gamma_{[\mu_1\cdots\mu_5|}\epsilon_2)f^{IJK}A_{|\mu_6}{}^J A_{\mu_7]}{}^K$ in $[\delta_{Q_1}, \delta_{Q_2}]\, B_{[7]}{}^I$ *would not* be absorbed into any gauge transformation, and thus the supersymmetry closure *would be* inconsistent. So, $K_{[6]}$ should *not* be entirely gauged away, maintaining supersymmetry closure. The non-trivial transformation $\delta_K B_{[7]} \neq 0$ is also closely related to this fact. In other words, if we *gauged away* $K_{[6]}$, the $\delta_K$-gauge freedom *would be* lost, and supersymmetry *would not* close. This is a typical example showing that even *non-physical* fields are playing important roles for the closure of supersymmetry.

As for $[\delta_{Q_1}, \delta_{Q_2}]\, C_{[8]}{}^I$, there arise three sorts of terms: $FB$, $\lambda^2$ and $K\lambda^2$-terms. The $\lambda^2$-terms need a special $\gamma$-matrix identities

$$(\gamma_{[\mu_1\cdots\mu_4|\nu})_{\alpha\beta}(\gamma_{|\mu_5\cdots\mu_8]}{}^\nu)_{\gamma\delta} = 0 \quad , \quad (\gamma_{[\mu_1\cdots\mu_6|}{}^{[3]})_{\alpha\beta}(\gamma_{|\mu_7\mu_8][3]})_{\gamma\delta} = 0 \quad . \tag{5.8}$$

Here all spinorial indices are for the negative chirality, contracted with positive chiral spinors, such as $\epsilon_1^\alpha$, $\epsilon_2^\beta$ or $\lambda^{\gamma\, I}$. Eq. (5.8) excludes possible $\gamma^{[5]}$ and $\gamma^{[9]}$-terms in the commutator.

Other non-vanishing commutators among $\delta_Q,\ \delta_T,\ \delta_U,\ \delta_K$ are

$$[\delta_Q,\ \delta_U] = \delta_{V_3} \quad , \qquad \gamma_{3\mu_1\cdots\mu_7}{}^I \equiv -7 f^{IJK} (\overline{\epsilon}\gamma_{[\mu_1}\lambda^J) \beta_{\mu_2\cdots\mu_7]}{}^K \quad , \tag{5.9a}$$

$$[\delta_Q,\ \delta_K] = \delta_{U_3} \quad , \qquad \beta_{3\mu_1\cdots\mu_6}{}^I \equiv -6 \kappa_{[\mu_1\cdots\mu_5} (\overline{\epsilon}\gamma_{\mu_6]}\lambda^I) \quad . \tag{5.9b}$$

Our supersymmetric EM-duality relationships are now

$$F_{\mu\nu}{}^I \overset{*}{=} + \frac{1}{8!} \epsilon_{\mu\nu}{}^{[8]} G_{[8]}{}^I \equiv \widetilde{G}_{\mu\nu}{}^I \quad , \tag{5.10a}$$

$$H_{[9]}{}^I \overset{*}{=} - \frac{1}{2} f^{IJK} (\overline{\lambda}^J \gamma_{[9]} \lambda^K) \quad , \tag{5.10b}$$

$$\lambda^I \overset{*}{=} - \chi^I \quad , \tag{5.10c}$$

$$\not{D}\lambda^I \doteq 0 \quad , \qquad \not{D}\chi^I \doteq 0 \quad , \tag{5.10d}$$

$$L_{[7]} \doteq 0 \quad . \tag{5.10e}$$

One difference compared with the previous 4D case is the new tensor $L_{[7]}$ needed for the supersymmetry-closure of the system. This will be mentioned in the next section.

## 6. 10D Superspace Re-Formulation

As re-confirmation and for future applications, we re-formulate the 10D result in superspace [19]. Our superfield-strengths are defined by

$$F_{AB}{}^I \equiv + E_{[A} A_{B)}{}^I - T_{AB}{}^C A_C{}^I + m f^{IJK} A_A{}^J A_B{}^K \quad , \tag{6.1a}$$

$$\begin{aligned} G_{A_1\cdots A_8}{}^I \equiv & + \frac{1}{7!} \nabla_{[A_1} B_{A_2\cdots A_8)}{}^I - \frac{1}{6!\cdot 2} T_{[A_1 A_2|}{}^C B_{C|A_3\cdots A_8)}{}^I \\ & + m C_{A_1\cdots A_8}{}^I - \frac{1}{6!\cdot 2} K_{[A_1\cdots A_6} F_{A_7 A_8)}{}^I \quad , \end{aligned} \tag{6.1b}$$

$$H_{A_1\cdots B_9}{}^I \equiv + \frac{1}{8!} \nabla_{[A_1} C_{A_2\cdots A_9)}{}^I - \frac{1}{7!\cdot 2} T_{[A_1 A_2|}{}^C C_{C|A_3\cdots A_9)}{}^I - \frac{1}{7!\cdot 2} f^{IJK} F_{[A_1 A_2}{}^J B_{A_3\cdots A_9)}{}^K , \tag{6.1c}$$

$$L_{A_1\cdots A_7} \equiv + \frac{1}{6!} E_{[A_1} K_{A_2\cdots A_7)} - \frac{1}{5!\cdot 2} T_{[A_1 A_2|}{}^B K_{B|A_3\cdots A_7)} \quad . \tag{6.1d}$$

In particular, the $KF$-term in (6.1b) is the superspace generalization of (5.2b) in component language.

These field-strengths satisfy the superspace BIds

$$\frac{1}{2} \nabla_{[A} F_{BC)}{}^I - \frac{1}{2} T_{[AB|}{}^D F_{D|C)}{}^I \equiv 0 \quad , \tag{6.2a}$$

$$\frac{1}{8!} \nabla_{[A_1} G_{A_2\cdots A_9)}{}^I - \frac{1}{7!\cdot 2} T_{[A_1 A_2|}{}^B G_{B|A_3\cdots A_9)}{}^I + \frac{1}{7!\cdot 2} L_{[A_1\cdots A_7} F_{A_8 A_9)}{}^I - m H_{A_1\cdots A_9}{}^I \equiv 0 \ , \tag{6.2b}$$

$$\frac{1}{9!} \nabla_{[A_1} H_{A_2\cdots A_{10})}{}^I - \frac{1}{8!\cdot 2} T_{[A_1 A_2|}{}^B H_{B|A_3\cdots A_{10})}{}^I - \frac{1}{8!\cdot 2} f^{IJK} F_{[A_1 A_2}{}^J G_{A_3\cdots A_{10})}{}^K \equiv 0 \quad , \tag{6.2c}$$

$$\frac{1}{7!} \nabla_{[A_1} L_{A_2\cdots A_8)} - \frac{1}{6!\cdot 2} T_{[A_1 A_2|}{}^B L_{B|A_3\cdots A_8)} \equiv 0 \quad . \tag{6.2d}$$

These are respectively referred to as $(ABC)_F$, $(A_1 \cdots A_9)_G$, $(A_1 \cdots A_{10})_H$ and $(A_1 \cdots A_8)_L$-BIds. Here $L_{A_1 \cdots A_7}$ plays an important role, as will be clarified shortly.

The superspace constraints at engineering dimensions $0 \le d \le 1$ are

$$T_{\alpha\beta}{}^c = +2(\gamma^c)_{\alpha\beta} \quad , \qquad L_{\alpha\beta c_1 \cdots c_5} = +2(\gamma_{c_1 \cdots 5})_{\alpha\beta} \quad , \tag{6.3a}$$

$$F_{\alpha b}{}^I = +(\gamma_b)_{\alpha\beta}\lambda^{\beta I} \equiv -(\gamma_b \lambda^I)_\alpha \quad , \qquad G_{\alpha b_1 \cdots b_7}{}^I = +(\gamma_{b_1 \cdots b_7})_{\alpha\beta}\chi^{\beta I} \equiv -(\gamma_{b_1 \cdots b_7}\chi^I)_\alpha \quad , \tag{6.3b}$$

$$\nabla_\alpha \lambda^{\beta I} = + \frac{1}{2} (\gamma^{cd})_\alpha{}^\beta F_{cd}{}^I \quad , \qquad \nabla_\alpha \chi^{\beta I} = + \frac{1}{8!} (\gamma^{cd})_\alpha{}^\beta G_{[8]}{}^I \quad . \tag{6.3c}$$

Here the upper (or lower) spinorial indices $^{\alpha,\ \beta,\ \cdots}$ (or $_{\alpha,\ \beta,\ \ldots}$) are for the positive (or negative) chiralities. We also use the collective indices $\underline{\alpha} \equiv ({}_\alpha, {}^\alpha)$, $\underline{\beta} \equiv ({}_\beta, {}^\beta)$, $\cdots$. Due to the mixed chirality $C_{\alpha\dot{\beta}}$ or $C^{\alpha\dot{\beta}}$ for the charge-conjugation matrices in 10D, the upper (or lower) indices are equivalent to *dotted* indices: $X^\alpha = C^{\alpha\dot{\beta}} X_{\dot{\beta}}$ (or $X_\alpha = -C_{\alpha\dot{\beta}} X^{\dot{\beta}}$). However, we avoid to use the dotted ones. All other independent components, such as $T_\alpha{}^{\beta c}$, $G_{\underline{\alpha}\underline{\beta} c_1 \cdots c_6}{}^I$, $H_{\underline{\alpha}\underline{\beta} c_1 \cdots c_7}{}^I$, *etc.* are zero.

The superspace constraints at $d = 3/2$ are

$$\nabla_\alpha F_{bc}{}^I = + (\gamma_{[b} \nabla_{c]} \lambda^I)_\alpha \quad , \qquad \nabla_\alpha G_{b_1 \cdots b_8}{}^I = + \frac{1}{7!} (\gamma_{[b_1 \cdots b_7} \nabla_{b_8]} \chi^I)_\alpha \quad , \tag{6.4a}$$

$$\nabla_\alpha H_{b_1 \cdots b_9}{}^I = - \frac{1}{2} f^{IJK} (\gamma^{cd} \gamma_{b_1 \cdots b_9} \lambda^J)_\alpha F_{cd}{}^K \quad . \tag{6.4b}$$

Our supersymmetric EM-duality relations are parallel to the component case (5.8):

$$F_{ab}{}^I \overset{*}{=} + \frac{1}{8!} \epsilon_{ab}{}^{[8]} G_{[8]}{}^I \equiv \widetilde{G}_{ab}{}^I \quad , \tag{6.5a}$$

$$G_{[8]}{}^I \overset{*}{=} - \frac{1}{2} \epsilon_{[8]}{}^{ab} F_{ab}{}^I \equiv -\widetilde{F}_{[8]}{}^I \quad , \tag{6.5b}$$

$$H_{[9]}{}^I \overset{*}{=} - \frac{1}{2} f^{IJK} (\overline{\lambda}^J \gamma_{[9]} \lambda^K) \quad , \tag{6.5c}$$

$$\lambda^{\alpha I} \overset{*}{=} - \chi^{\alpha I} \quad , \tag{6.5d}$$

$$(\nabla\!\!\!\!/\, \lambda^I)_\alpha \doteq 0 \quad , \qquad (\nabla\!\!\!\!/\, \chi^I)_\alpha \doteq 0 \quad , \tag{6.5e}$$

$$L_{a_1 \cdots a_7} \doteq 0 \quad . \tag{6.5f}$$

The satisfaction of the BIds (6.2) needs special care, in particular, the role played by the superfield-strength $L_{A_1 \cdots A_7}$. For example, if the $LF$-term in (6.2b) *did not* exist in the $(\alpha\beta\gamma d_1 \cdots d_6)_G$-BId at $d = 1/2$, then a term proportional to $(\gamma_{[d_1|})_{(\alpha\beta|} (\gamma_{|d_2 \cdots d_6]} \chi^I)_{|\gamma)}$ *would* be left over. This term is cancelled by the like-term arising from the $LF$-term in the G-BId (6.2b). Similarly at $d = 1$, the $(\alpha\beta c_1 \cdots c_7)_G$-BId, which is equivalent to the closure $[\delta_{Q_1}, \delta_{Q_2}] B_{[7]}{}^I$ in component language, works as follows: If there *were no* $LF$-term in this BId, then there *would remain* a term $-(1/6!)(\gamma_{[c_1 c_2|}{}^{[3]})_{\alpha\beta}\, G_{[3]|c_3 \cdots c_7]}{}^I \overset{*}{=} -(1/120)(\gamma_{[c_1 \cdots c_5|})_{\alpha\beta} F_{|c_6 c_7]}{}^I$, upon the use of the duality (5.10a). However, this term is exactly cancelled by the like-term arising from $(1/240) L_{\alpha\beta [c_1 \cdots c_5|} F_{|c_6 c_7]}{}^I$. We have thus confirmed the significance of the $K_{[6]}$-field both in component and superspace languages. The significance of the $\delta_K$ for the closure of supersymmetry in component is reflected into the necessity of the $LF$-term in $G$-Bianchi identity (6.2b) in superspace.

For BIds at $d = 1/2$, the following $\gamma$-matrix relationships are crucial:

$$(\gamma_e)_{(\alpha\beta}(\gamma^{e f_1 \cdots f_4})_{\gamma\delta)} = 0 \quad , \tag{6.6a}$$

$$(\gamma_{[a|}{}^{[4]})_{\alpha\beta}(\gamma_{|b][4]})_{\gamma\delta} = 0 \quad , \tag{6.6b}$$

$$(\gamma^{[e_1})_{(\alpha\beta}(\gamma^{e_2 \cdots e_6]})_{\beta\delta)} = 0 \quad , \tag{6.6c}$$

in addition to (5.8). All of these can be easily confirmed by the use of more fundamental relationships, such as

$$\delta_{(\alpha}{}^\gamma \delta_{\beta)}{}^\delta = -\frac{1}{8}\,(\gamma_e)_{\alpha\beta}(\gamma^e)^{\gamma\delta} - \frac{1}{1920}\,(\gamma_{[5]})_{\alpha\beta}(\gamma^{[5]})^{\gamma\delta} \quad . \tag{6.7}$$

As in the 4D case in section 3, we can confirm the internal consistency of supersymmetric EM-duality in (6.5). A typical example is the spinorial derivative $\nabla_\alpha$ acting on (6.5a) or (6.5c), yielding zero by the use of other duality-related equations in (6.5). These are parallel to the component case, so that we do not give details.

## 7. Concluding Remarks.

In this paper, we have accomplished the $N = 1$ supersymmetrization of the EM-duality relationship (2.6) for *non-Abelian* gauge groups in 4D. The original EM-duality (2.6) is supersymmetrized to the equations in (3.3). Subsequently, we have also established the EM-duality (5.10) for $N = (1, 0)$ non-Abelian supersymmetric system in 10D.

The total system in 4D is simple with only three multiplets: a YMVM, a DVM and a *non-physical* TM. Yet the TM plays a very crucial role for avoiding the conventional problem with non-Abelian EM-duality based on tensor-hierarchy [12][13][14]. Even though EM-duality for $SO(8)$ group with $N = 8$ *local* supersymmetry [22] had been known for a long time, our system is simple only with *global* supersymmetry. Our formulations became possible, thanks to the recently-developed tensor-hierarchy formulation [12][13][14].

We have confirmed the total consistency both in component and superspace languages [19] both in 4D and 10D, as well. The existence of the extra tensors, such as $C_{\mu\nu}{}^I$ in 4D or $C_{[8]}{}^I$ and $K_{[6]}$ in 10D is to maintain the total consistency of the system. In particular, the field-strengths $G$ and $H$ contain CS-like terms, guaranteeing consistency. This aspect is also the result of tensor-hierarchy formulation [12][13][14].

The validity of the particular $KF$-type CS-term in the $G$-field strength (5.2b), and the $LF$-term in the $G$-Bianchi identity (5.3b) in component language is re-confirmed as (6.1b) and (6.2b) in superspace. The necessity of the potential $K_{[6]}$ or its field strength $L_{[7]}$ is confirmed both in component and superspace languages. It is the sophisticated combination of tensor-hierarchy formalism [13][14] and the special role played by $K_{[6]}$ and $L_{[7]}$ that make our EM-duality possible in 10D.

In our paper, we have dealt with the manifestly-Lorentz-covariant EM-duality, such as $F_{\mu\nu}{}^I \stackrel{*}{=} + (1/8!)\, \epsilon_{\mu\nu}{}^{[8]} G_{[8]}{}^I$ in 10D, instead of *non-manifest* Lorentz covariance as in [3]. Even though our system lacks a lagrangian formulation, it still maintains *manifest* Lorentz-covariance at the field-equation level.

As some readers may have noticed, (3.3d) indicates that the dual field-strength $\widetilde{H}_\mu{}^I$ equals the YM-current vector: $\widetilde{H}_\mu{}^I \stackrel{*}{=} J_\mu{}^I$. The divergence of the LHS of this relationship vanishes by the EM-duality (2.6) *via* the $H$-BId (2.2c), while the vanishing of the RHS is the usual current conservation. In other words, the new relationship like (3.3d) relates the current $J_\mu{}^I$ directly to field-strength $\widetilde{H}_\mu{}^I$ *without* involving derivatives of the latter.

We believe our present result may well be important for generating other and new supersymmetric consistent theories of non-Abelian vectors and tensors associated with general EM-dualities, in diverse space-time dimensions.